# Towards automation of computing fabrics using tools from the fabric management workpackage of the EU DataGrid project

Olof Bärring, Maite Barroso Lopez, German Cancio, Sylvain Chapeland, Lionel Cons, Piotr Poznañski, Philippe Defert, Jan Iven, Thorsten Kleinwort, Bernd Panzer-Steindel, Jaroslaw Polok, Catherine Rafflin, Alan Silverman, Tim Smith, Jan Van Eldik
*CERN, CH1211 Geneva-23, Switzerland*

Massimo Biasotto, Cristine Aiftimiei, Enrico Ferro, Gaetano Maron
*INFN-LNL, Viale dell'Universita 2, I-35020 Legnaro (PADOVA), Italy*

Andrea Chierici, Luca Dellagnello
*INFN-CNAF, Viale Berti Pichat 6/2, I-40127 Bologna, Italy*

Marco Serra
*INFN-Roma1, P.le Aldo Moro 2, I-00185 Roma, Italy*

Michele Michelotto
*INFN-PADOVA, Via Marzolo 8, I-35131 Padova, Italy*

Thomas Röblitz, Florian Schintke
*ZIB, Takustraße 7, D-14195 Berlin – Dahlem, Germany*

Lord Hess, Volker Lindenstruth, Frank Pister, Timm Morten Steinbeck
*Im Neuenheimer Feld 227, D-69120 Heidelberg, Germany*

David Groep, Martijn Steenbakkers
*NIKHEF, PO Box 41882, 1009 DB Amsterdam, The Netherlands*

Paul Anderson, Tim Colles, Alexander Holt, Alastair Scobie
*University of Edinburgh, Old College, South Bridge, Edinburgh EH8 9YL, UK*

Michael George
*Oxford Street, Liverpool L69 7ZE, United Kingdom*

Rafael A. García Leiva
*Departament of Theoretical Physics, Universidad Autónoma de Madrid
Ctra Colmenar Km 15 28049 Madrid, Spain*

The EU DataGrid project workpackage 4 has as an objective to provide the necessary tools for automating the management of medium size to very large computing fabrics. At the end of the second project year subsystems for centralized configuration management (presented at LISA'02) and performance/exception monitoring have been delivered. This will soon be augmented with a subsystem for node installation and service configuration, which is based on existing widely used standards where available (e.g. rpm, kickstart, init.d scripts) and clean interfaces to OS dependent components (e.g. base installation and service management). The three subsystems together allow for centralized management of very large computer farms. Finally, a fault tolerance system is being developed for tying together the above subsystems to form a complete framework for automated enterprise computing management by 3Q03. All software developed is open source covered by the EU DataGrid project license agreements.

This article describes the architecture behind the designed fabric management system and the status of the different developments. It also covers the experience with an existing tool for automated configuration and installation that have been adapted and used from the beginning to manage the EU DataGrid testbed, which is now used for LHC data challenges.

## 1. INTRODUCTION

The EU DataGrid project is a three-year EU funded project to develop middleware for data intensive grid applications. The project started in January 2001 and is divided into 12 workpackages: 5 for middleware development, 2 for Grid testbed infrastructure, 3 for scientific applications (HEP, Biology, Earth Science) and 2 for dissemination and management.

Workpackage 4, WP4, is one of the middleware workpackages and has as main objectives to deliver a computing fabric comprised of all the necessary tools to manage a center providing grid services on clusters of thousands of nodes. The workpackage is divided into five software development subtasks:

- Configuration management
- System monitoring
- Installation management and maintenance
- Fault tolerance
- Resource management
- "Gridification"

The four first subtasks provide the basic software subsystems for the automated fabric management while the two latter are more aimed to grid-enable the fabric. This paper will mainly describe the automated fabric management subsystems.

The next section describes the high-level architecture for how the different subsystems work together. Thereafter follows four sections with detailed descriptions for each subsystem and the development status. Finally the

**MODT004**

conclusions will sum up the experience and status so far and future work up to the end of the project.

## 2. ARCHITECTURE

A high level view of the WP4 architecture for automated fabric management is depicted in Figure 1.

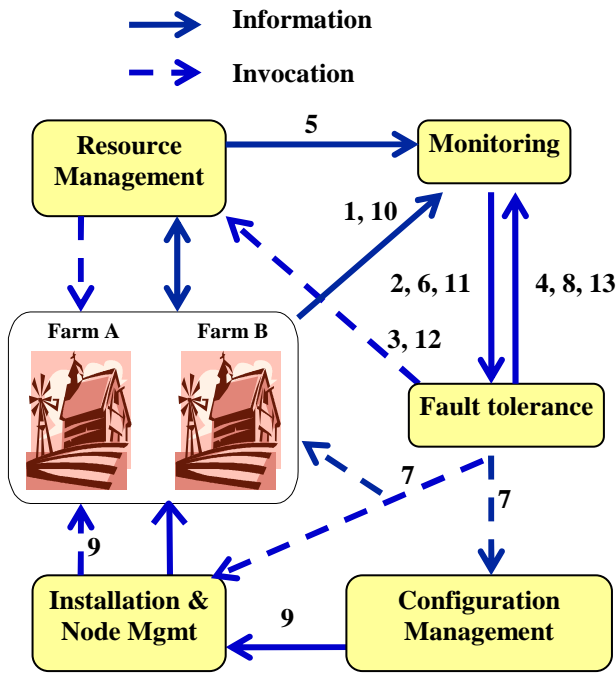

**Figure 1: high level view of the WP4 architecture for automated fabric management.**

The numbered arrows 1 – 10 indicated in Figure 1 describe a typical sequence for the detection of an exception and coordination of the automated recovery:

1. Monitoring data flows from all nodes in the computer center to a monitoring repository
2. The fault tolerance system reads and correlates data from the monitoring system to detects exception conditions that require automated interventions
3. The fault tolerance system instructs the resource management system that interfaces with the cluster batch system to remove the node from production and drain/kill (depending on the urgency) running jobs on the node
4. Fault tolerance system informs the monitoring system about the action taken to remove the node from production
5. The resource management system informs the monitoring when the node has been removed from production and all running jobs have finished
6. The fault tolerance system reads and the updated node information and decides to launch the node maintenance
7. Depending on the maintenance action the fault tolerance system may first change the configuration of the node by committing a new configuration template in the configuration management system. If no reconfiguration is required the fault tolerance system may either directly launch the repair on the node or instruct the installation and node mgmt system to perform some action (e.g. repair an installation or simply reboot the node).
8. Fault tolerance system informs the monitoring system about the launched repair action
9. Installation and node management system reads the node configuration profile managed by the configuration management system and calls service configuration objects to deploy new configurations or simply restart services.
10. Monitoring data flows from all nodes in the computer center to a monitoring repository
11. Fault tolerance system reads data from the monitoring repository and detects that the node has been repaired.
12. Fault tolerance system instructs the resource management system to put back the node in production
13. Fault tolerance system informs the monitoring system about the action to put the node back in production

It is important to point out the information model chosen by WP4: the configuration management system manages the *desired* state while the monitoring system records the *actual* state. This distinction is necessary since deployment of configuration changes may take long time and in real production clusters the actual state and desired state may only converge asymptotically. For instance, a configuration change that requires a reboot cannot be deployed on all nodes at the same time if a minimum service level is required.

The monitoring system not only receives normal monitoring data such as performance and exception metrics but it is also used to keep track of all automatic recovery actions. It works in conjunction with the fault tolerance system, which takes its input data from the monitoring system and reports back the launched recovery actions. This strict tracing of actions also holds true for manual interventions where, for instance, the acknowledgement of an alarm is recorded by the monitoring system. The aim is to allow for several levels of fault tolerance recovery so that if a repair fails after a certain number of retries, another repair strategy could be selected automatically.

The configuration information for the desired state is expressed in a special declarative language, called the High Level Definition Language, HLDL, developed by WP4. The administrators or service managers write HLDL configuration *templates*, which are compiled by the configuration management system into node profiles. A node profile is an XML files containing the entire configuration that is to be managed on a node. The HLDL language supports inheritance through inclusion, which allows for managing the configuration information in a hierarchical structure called the template hierarchy. Another hierarchy, the configuration schema, is formed by the name space defined for the configuration parameters. The configuration schema and the template hierarchy are independent.

The installation and node management subsystem includes several components:
- The Automated Installation Infrastructure (AII) for automatic generation of DHCP configuration and kickstart files according to the desired configuration managed by the configuration management system.
- The Node Configuration Manager (NCM) deploys the desired node configuration using a component framework.
- The Software Package Management Agent (SPMA) handles local software installation.
- The software repository (SWRep) contains the software packages that might be referenced from a desired configuration.

In the following sections the monitoring, fault tolerance, configuration and installation subsystems will be described in more detail.

## 3. SYSTEM MONITORING SUBSYSTEM

### 3.1. Design

The different components of the monitoring subsystem [13] are shown in Figure 2. A Monitoring Sensor Agent (MSA) runs on all monitored nodes. It is responsible for calling the plug-in sensors to sample the configured metrics. The sampling frequency can be configured per metric. The interface is designed so that the sensor is not required to answer to sampling requests and it may chose to trigger its own unsolicited samplings. The sensor communicates with the MSA over a normal UNIX using a simple text protocol. To hide the protocol details, a sensor API C++ class has been defined for convenience.

All monitoring data gathered on a node is stored in a local cache, which is available for local consumers of monitoring data. This is useful for allowing for local fault tolerance correlations engines. The monitoring data is also forwarded to a global measurement repository, where it is available for remote global consumers. The same externalized measurement repository API is used to access the data at both local and global level. The repository API is implemented using SOAP RPC and provides methods for time series queries and subscription/notification of new monitoring measurements. The sampling values are plain text strings and it is up to the consumer to correctly parse the values. While this can be perceived as a cumbersome for simple single number valued metrics, it has the advantage that the metric values are unconstrained as long as they can be represented as printable text strings.

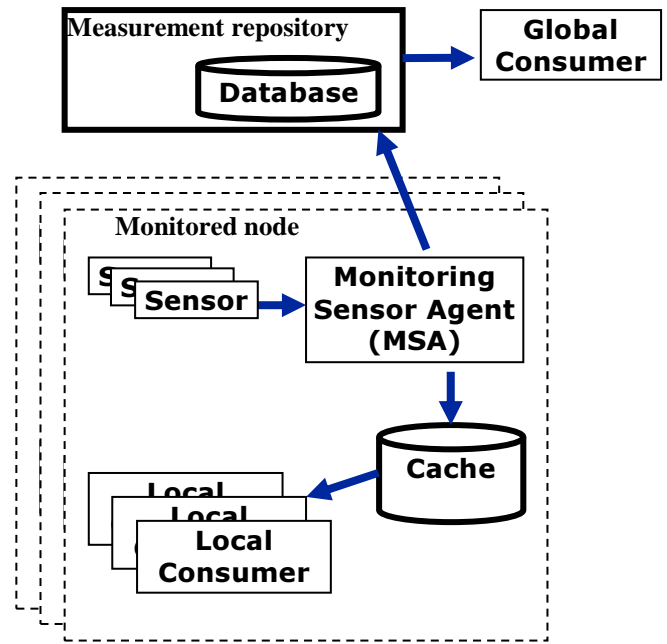

**Figure 2: components of the monitoring system and how they are deployed**

The local cache is implemented as a flat text file database, with one file per metric per day. The file format is "timestamp value". The global measurement repository server provides an open interface (same as the repository API) to plug-in any backend database system. Current database backend implementations for flat text file (same as for the local cache) and Oracle exist. An interface MySQL is being developed.

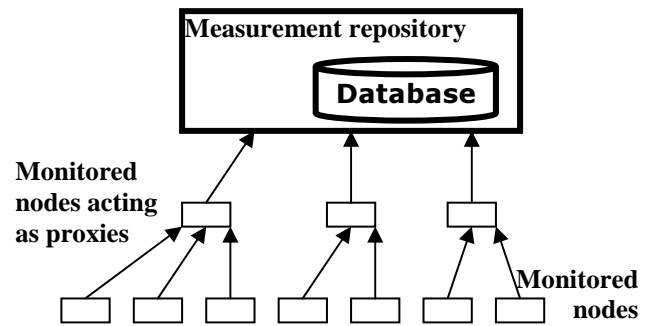

**Figure 3: schematic picture of TCP transport proxy mechanism.**

The transport of monitoring data from the monitored nodes to the central repository is also pluggable. An UDP based implementation has been in use since more than one year at CERN. A TCP based implementation exists as prototype. The TCP based solution works over permanently open sockets and it includes a proxy like mechanism to fan-out the number of open connections on the global repository to a subset of the monitored nodes.

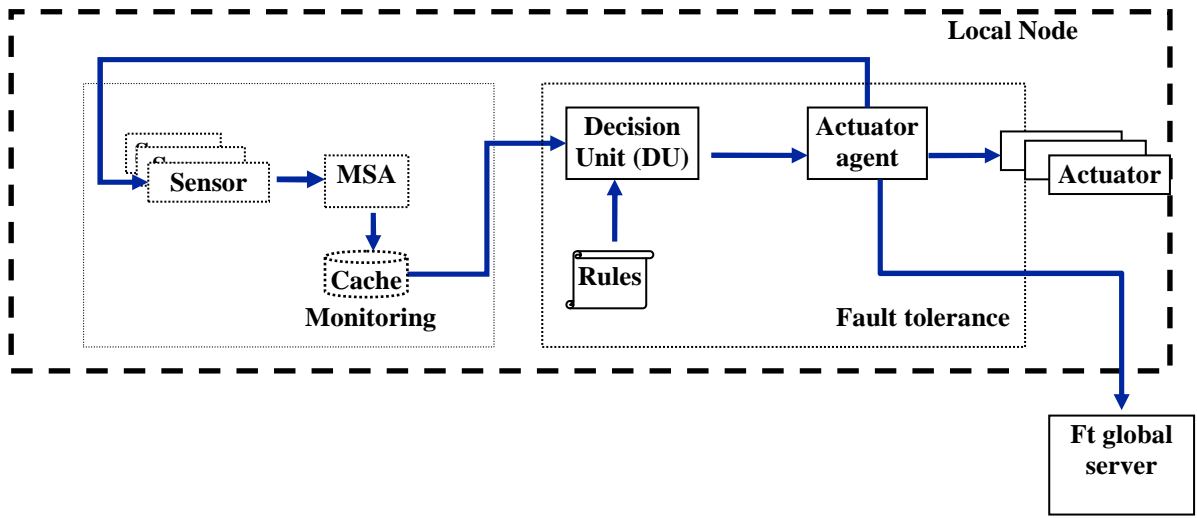

**Figure 4: fault tolerance subsystem and its interaction with the monitoring subsystem**

On the "proxy" nodes the transport part of the MSA not only sends the monitoring data of the node itself but it also receives and forwards data from other MSAs. The proxy architecture is schematically shown in Figure 3. Note that instead of 10 incoming connections to the global measurement repository there are only 3 for the configuration depicted in the figure. Currently the proxy configuration must be configured statically but it is intended to make this dynamic so that the measurement repository decides on-the-fly which of its clients should act as proxies.

### 3.2. Status

Monitored nodes:
- Monitoring Sensor Agent (MSA) and UDP based transport protocol are ready and used on CERN production clusters since more than a year
- The TCP based proprietary protocol exists as prototype. More testing and functionality needed to be ready for production use

Central services
- Repository server exists with both flatfiles and Oracle database. The latter is currently being evaluated for production use at CERN. Support for MySQL is planned for later in 2003
- Alarm display: still in early prototype phase.

Repository API for local and global consumers:
- C library implementation of API (same for local and global consumers)
- Bindings for other languages can be generated directly from the WSDL

## 4. FAULT TOLERANCE SUBSYSTEM

### 4.1. Design

The components of the fault tolerance subsystem and how they interoperate with the monitoring subsystem are shown in Figure 4. Central to the fault tolerance system is the rule based correlation engine allowing users/administrators to define set of rules that are executed by the system. A rule determines exception conditions and maps them into actions to be executed. Arbitrarily complex exception conditions can be expressed using a simple but efficient language. The language offers the basic numerical and Boolean operations as well as string comparison. The language also provides a possibility to collect all kinds of data from a computing node, which then enters into the expression as a variable.

A web-based XML editor can be used for creating the rules. The XML file defining the rule is copy to the nodes where it is parsed by a local fault tolerance daemon consisting of a decision unit and an actuator agent.

The decision unit parses the configured rules and is responsible for using the monitoring repository API to call the monitoring system to subscribe to all metrics needed by the rule. The monitoring system notifies the decision unit whenever there is a new measurement of the metrics. The decision unit then re-evaluates the rule. The actuator agent is called if an exception condition expressed by the rule is met.

The actuator agent takes the output from the decision unit and determines which actuator to call. An actuator can be any executable command (binary or script) that is available on the node. The actuator agent launches the actuator and reports back as a normal metric the return status of the actuator to the monitoring system. This feedback to the monitoring system is important. It allows tracing actions and it allows the correlation engine to be state less. Retry loops can be created by defining a rule that takes the actuator return status metric as input. The feedback also allows for escalation of exception that cannot be solved locally.

### 4.2. Status

The fault tolerance subsystem is not yet ready for production deployment but a prototype was demonstrated

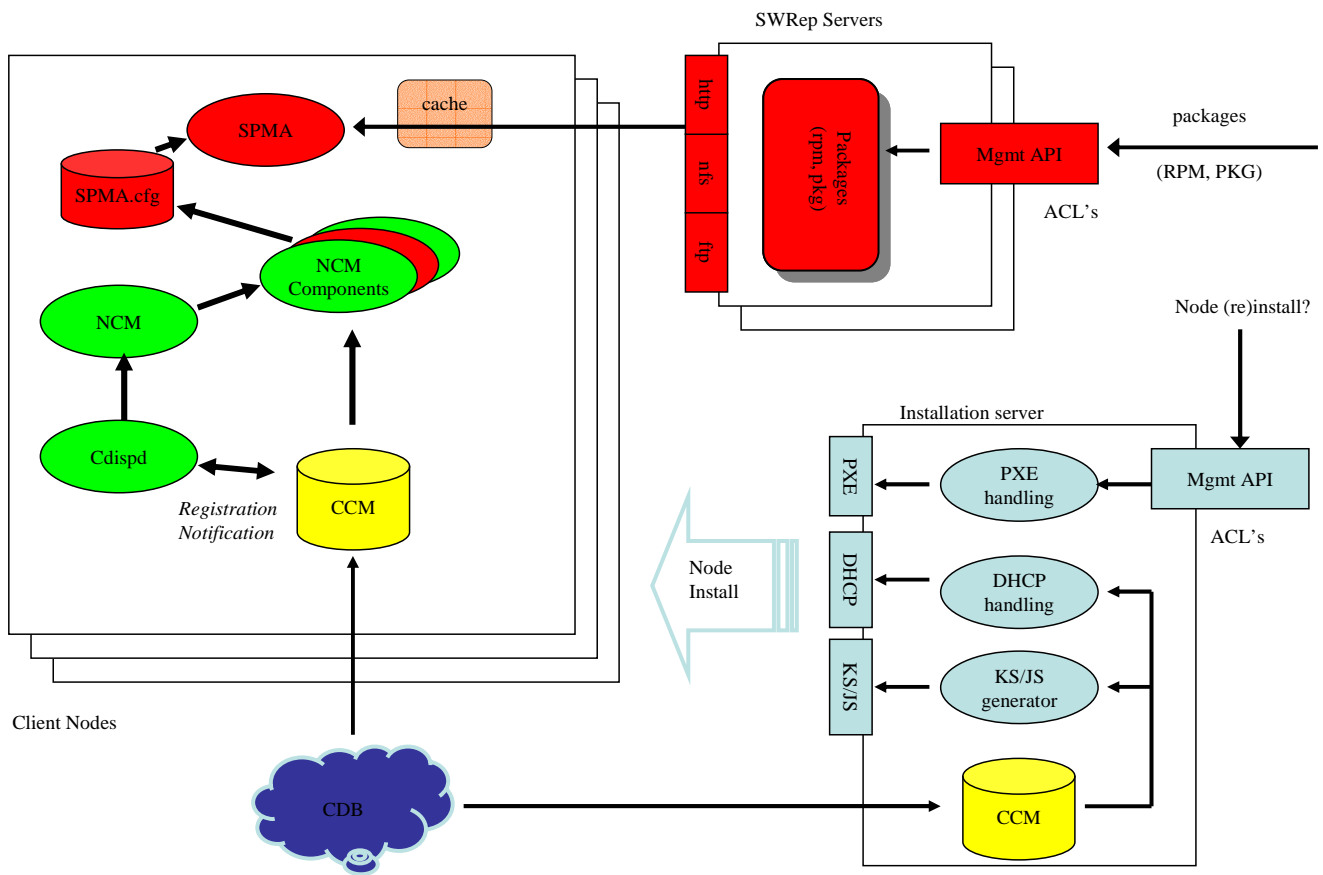

**Figure 5:** Installation Management subsystem

working together with the fabric monitoring system at EU review in February 2003. The setup included a Web-based rule editor and central rule repository (MySQL) for managing the rules. On the local nodes a local fault tolerance daemon was deployed that
- Automatically subscribed to monitoring metrics specified by the rules
- Launched the associated actuators when the correlation evaluates to an exception
- Reported back to the monitoring system the recovery actions taken and their status

Only local correlations (detection of daemon dead followed by an automatic restart) were demonstrated at the review. Global (inter-node) correlations will be supported later.

## 5. INSTALLATION MANAGEMENT SUBSYSTEM

The installation management system ([3]) provides scalable solutions for the automated from scratch installation, (re-)configuration and software package distribution and management of large clusters.

### 5.1. Automated Installation Infrastructure

The AII (Automated Installation Infrastructure) subsystem [4] provides tools for the management of standard vendor installation servers. This includes the configuration of network related information, like the DHCP tables and the network bootstrap protocol (e.g. PXE for Intel/Linux and OpenBoot for Solaris). Also, the node specific installation setup rules (KickStart for Linux respective JumpStart for Solaris) have to be generated. The AII obtains its configuration either from the CDB (via the CCM), or from site specific network databases.

### 5.2. Node Configuration Management

The NCM (Node Configuration Management) subsystem [5] provides a framework for adapting the *actual* configuration of a node to its *desired* configuration, as it is described in the node's profile inside the CDB.

Plug-in software modules called 'components' are responsible for the configuration of local services (e.g. network, sendmail, NFS), analogously to LCFG 'objects' [6] or SUE 'features' [7]. For this, they can read CDB configuration information via the CCM, and create/update/delete local service configuration files in order to match the CDB configuration description. Components register which configuration entries or subtrees it is interested in, and get notified in case of changes.

Each component contains the knowledge for translating the CDB configuration into each local service's specific config file syntax. A component may also require notifying a service about a configuration change (e.g. by running a 'restart' or 'reload' method in a SysV init script).

The NCM subsystem contains the following modules:

- **cdispd**: The Configuration Dispatch Daemon (cdispd) monitors the node configuration profile by polling the CCM. In case of changes in the configuration profile, the cdispd will invoke the affected components via the ncd.
- **ncd**: The Node Configuration Deployer (ncd) is the framework and front-end for executing the configuration components. The ncd can be executed manually, via cron, or via the cdispd. It takes care of which takes care of configuration locking and inter-component dependency ordering prior to executing components sequentially.
- **Component support libraries**: Libraries for recurring system management tasks (system information, interfaces to system services, file editing), log file handling, interface to Monitoring, etc.

### 5.3. Software package management and distribution

The Software Package Management and Distribution (SPM) subsystem [8] is responsible for managing and storing software packages, and the distribution and installation of these packages on client nodes.

The SPM subsystem contains the following modules:

- **Software Repository**: The Software Repository (SWRep) module allows site administrators and package maintainers to store and manage software packages (like RPM or PKG packages) subject to authentication and authorization using ACL's. The packages themselves are accessible to the clients via standard protocols including HTTP, FTP, or using a shared file system. It is possible to have multiple (replicated or independent) Software Repository instances for a given fabric, allowing for load balancing, and also private per-department repositories. The replication of repositories can be done with standard tools like rsync.
- **SPMA**: The Software Package Manager Agent (SPMA) runs on the target nodes. It reads a local configuration file with the list of desired packages, compares it with the currently installed packages, computes the necessary install/deinstall/upgrade operations, and invokes the system packager (e.g. rpm[1] on Linux, pkgadd/del on Solaris) with the right operation transaction set.
- **SPM component**: The information on which packages are to be deployed on which nodes (desired or target configuration), and which packages are available on which repositories can be kept in the CDB. The SPM component fits into the NCM framework described above. It retrieves the list of packages to be installed for the current node from the CDB via the CCM, creates with this information a local configuration file for the SPMA, and launches the SPMA.

Typically, the SPMA is used for managing all packages on a node. This is useful for nodes which are under full control of the fabric management system. However, for add-on installations or desktop systems, the SPMA can be run in 'light' mode, taking care of a subset of packages only, according to configurable policies.

For performance and scalability issues, the SPMA can use a local cache where packages can be stored ahead. This way, peak loads on software repository servers can be avoided during upgrades of large farms, but keeping consistency across the upgraded nodes. Also, the default transport protocol is set to HTTP for its scalability and low overhead.

### 5.4. Status

The architectural design of the AII and NCM subsystems has finished; the detailed design and implementation is progressing, and a first prototype version of these subsystems will be available at the end of the summer. A integrated solution, including components for configuring the most common system services, is expected to be available by the end of September.

A first production version of the SPM subsystem is being deployed on CERN's central batch and interactive services.

## 6. CONFIGURATION MANAGEMENT SUBSYSTEM

### 6.1. Design

The Configuration Management subsystem consists of modules shown in the Figure 6. The configuration information is stored centrally in the Configuration Database, CDB. The configuration of a particular node is stored locally by the Configuration Cache Manager and available to the clients through Node View Access API.

---

[1] Since 'rpm' on Linux does not accept multiple simultaneous operation types, we developed a new front-end called 'rpmt' (for transactional rpm) capable to handle multiple operations on multiple packages in a single transaction.

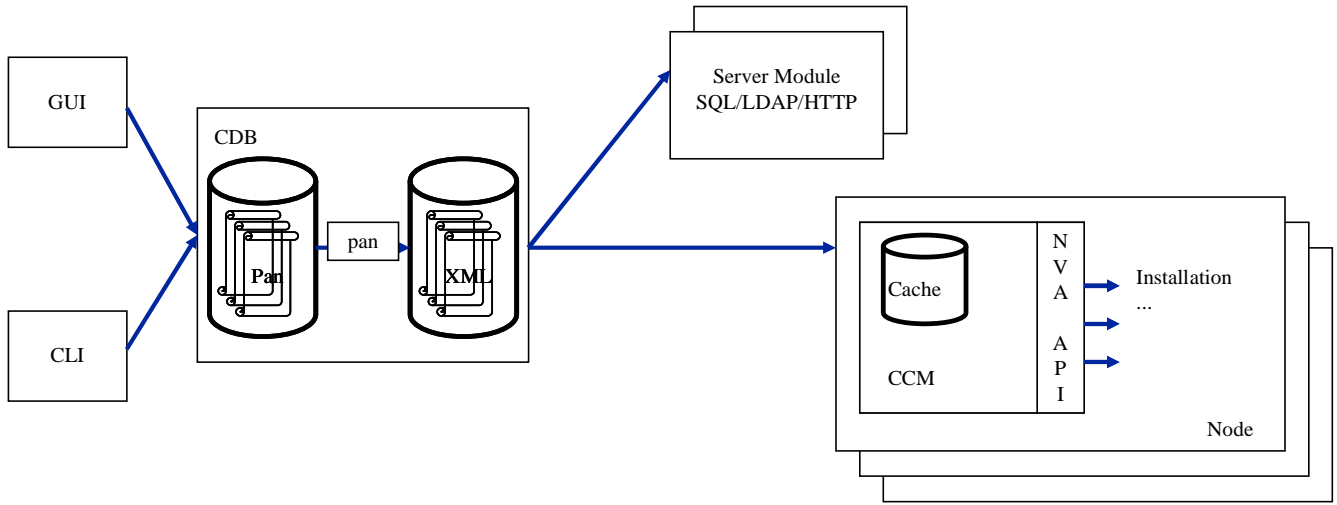

**Figure 6: Configuration Management subsystem**

Configuration Information may be also accessed centrally through some other means e.g. the SQL or LDAP queries. This functionality is provided by the Server Modules.

The configuration information is structured in a tree format, and expressed with the High Level Description Language called Pan [9]. Pan mainly consists of statements to set some value to a configuration parameter identified by its path in the configuration information tree.

Pan features include other statements like "include" (very similar to cpp's #include directive) or "delete" that removes a part of the configuration information tree.

The grouping of statements into templates allows the sharing of common information and provides simple inheritance mechanism.

Pan contains a very flexible typing mechanism. It has several built-in types (such as "boolean", "string", "long" and "double") and allows compound types to be built on top of these. Once the type of the element is known, the compiler makes sure that only values of the right type are assigned to it.

To have even greater control on the information generated by the compiler, one can attach validation code to a type or to a configuration path.

The validation code is represented in a simple yet powerful data manipulation language which is a subset of Pan and syntactically similar to C or Perl.

The Configuration Database [10] stores two forms of configuration information. One is High Level Description expressed in the Pan language. The other is the Low Level Description [11] and is expressed in XML.

The system administrators can edit the High Level Description, either through Command Line Interface (CLI) or Graphic User Interface (GUI). There is also possibility of having some scripting layer on top of the Configuration Database. The Low Level Description (one XML file per machine) is always generated using the Pan compiler.

The database works in a transactional way. It performs validation of the configuration information. Once the validation and compilation process is accomplished successfully, the changes introduced by the user are stored in the database and visible to its clients.

Configuration Database also provides mechanisms for versioning and it maintains the history of the changes of the configuration information.

The database itself includes a scalable distribution mechanism for the XML files based on HTTP, and the possibility of adding any number of back-ends (such as LDAP or SQL) - the Server Modules, to support various query patterns on the information stored. It should scale to millions of configuration parameters.

Configuration Cache Manager runs on every node and caches the XML machine configuration (to support disconnected operations). The access to the information is provided through a Node View Access API [12] that hides the details such as the XML schema used.

The Manager may poll the Configuration Database for the configuration information. It also receives UDP notification sent by the database if the machine's configuration is changed.

## 6.2. Status

The Configuration Management subsystem is implemented except for the Command Line Interface and the Server Modules. Most of the components are in production versions. It has been being deployed for the LCG1 using the "PanGUIn" GUI.

In parallel, the whole system is being consolidated. The issues of scalability and security are being studied and addressed. Currently the Server Modules for the XML replication and the SQL access are developed.

## 7. CONCLUSIONS AND OUTLOOK

In the first two years of the project the work was focused into surveying the existing solutions for automated management of large clusters, getting the architecture and design right, and implementing prototypes of the different subsystems.

At the time of writing this paper, stable prototypes exist for all subsystems; some of them are already deployed at CERN and/or the EDG project testbed, and are presently being evaluated for production purposes:

- System monitoring: 1000 nodes being monitored, with 20 different MSA configurations (form 80 to 120 metrics per node), sending data to a central measurement repository.
- Configuration management: CDB in production status, holding site-wide, cluster and node specific configurations for 550 clients, totalizing 1200 templates.
- Installation management and maintenance: First pilot of Software Repository and SPMA being deployed on CERN Computer Centre for the central CERN production (batch & interactive) services. All the nodes declared in the CDB are (re)installed using the SPMA, accessing packages from a replicated and load-balanced HTTP-based SWRep repository server cluster. The idea is go grow up to ~ 1.5K nodes at the end of 2003.

Consult reference [2] for further details.

Most of the changes required to move from the prototyping stage to production quality tools are the result of the testing and evaluation period. No fundamental flaw has been found in the architecture so far. However, simulating real production use has shown to be the only efficient way of finding development bugs or functionality enhancements. Computing Centers have also very high requirements on stability and reliability that can only be tested in a real environment.

The users have been involved throughout the whole design and development process, but it is now when their collaboration is becoming crucial.

The plans from now till the end of the project are focused on two areas. Firstly the work on general aspects as security, scalability, usability, graphical user interfaces, etc needs to be completed.

Secondly, the different fabric subsystems need to be "glued" together to build a consistent set of fabric management tools. Each subsystem has been implemented as a modular set of tools, which could be used independently according to the user needs. The tools will work together to provide all the functionality needed to automatically manage medium size to very large computing fabrics, as stated in the initial objectives.

## 8. ACKNOWLEDGMENTS

The authors wish to thank the EU and our national funding agencies for their support of this work.

.